\documentclass{PoS}
\usepackage{amsmath, amssymb}
\usepackage{graphicx}

\title{Exploring the QCD phase diagram at finite density by the complex Langevin method on a $16^3 \times 32$ lattice}

\ShortTitle{Exploring the QCD phase diagram at finite density...}

\author{\speaker{Shoichiro Tsutsui}\thanks{RIKEN-QHP-432, KEK-TH/2174}\\
		Theoretical Research Division, Nishina Center, RIKEN,
		Wako, Saitama 351-0198, Japan\\
        E-mail: \email{shoichiro.tsutsui@riken.jp}}

\author{Yuta Ito\\
        National Institute of Technology, Tokuyama College, Gakuendai, Shunan,
        Yamaguchi 745-8585, Japan \\
	E-mail: \email{y-itou@tokuyama.ac.jp}}

\author{Hideo Matsufuru\\
	KEK, 1-1 Oho, Tsukuba, Ibaraki 305-0801, Japan\\
	E-mail: \email{hideo.matsufuru@kek.jp}}

\author{Jun Nishimura\\
        KEK and Graduate University for Advanced Studies (SOKENDAI),\\
        1-1 Oho, Tsukuba, Ibaraki 305-0801, Japan\\
	E-mail: \email{jnishi@post.kek.jp}}

\author{Shinji Shimasaki\\
	Research and Education Center for Natural Sciences,
	Keio University, Hiyoshi 4-1-1, Yokohama, Kanagawa 223-85
	21, Japan \\
	E-mail: \email{shinji.shimasaki@keio.jp}}

\author{Asato Tsuchiya\\
	Department of Physics, Shizuoka University,
	836 Ohya, Suruga-ku, Shizuoka 422-8529, Japan \\
	E-mail: \email{tsuchiya.asato@shizuoka.ac.jp}}

\abstract{We explore the QCD phase diagram at finite density with four-flavor staggered fermions 
	using the complex Langevin method, which is a promising approach to overcome the sign problem. 
In our previous work
on an $8^3\times16$ lattice at $\beta=5.7$ with the quark mass $m= 0.01$, 
	we have found that the baryon number density has a clear plateau as a function of the chemical potential. 
	In this study, we use a $16^3\times32$ lattice to reduce finite volume effects and find that the plateau structure survives. Moreover, the number of quarks in the plateau region turns out to be 24, which is exactly the same as the one obtained previously on the $8^3\times16$ lattice. 
We provide a simple interpretation of this number, which suggests 
that the Fermi sphere is starting to form.
}

\FullConference{37th International Symposium on Lattice Field Theory - Lattice2019\\
		16-22 June 2019\\
		Wuhan, China}
	
\begin{document}
\section{Introduction}
One of the long-standing problems in QCD is 
to explore its phase diagram at finite density,
which is important in understanding the results 
of the heavy-ion collision experiments
and the interior structure of neutron stars.
However, nonperturbative aspects of QCD at finite density 
remain elusive 
due to the sign problem,
which makes conventional Monte Carlo methods inapplicable.

One of the promising approaches to overcome the sign problem is 
the complex Langevin method (CLM)~\cite{Klauder:1983sp,Parisi:1984cs}.
In this method, we consider a fictitious time evolution of 
the complexified dynamical variables
described by the Langevin equation
with the drift term given by the gradient of the complex action.
If this time evolution reaches a unique equilibrium,
physical quantities can be computed as expectation values 
by extending them holomorphically to functions of the complexified variables.
This procedure does not rely on the probabilistic interpretation 
of the path integral weight,
and hence the sign problem can be circumvented.
However, the equivalence between the CLM and the path integral formulation 
is highly nontrivial, and 
in fact it is known that the method does not always give correct results.

Recently, 
the condition for justifying the CLM has been 
clarified~\cite{Aarts:2009uq,Aarts:2011ax,Nishimura:2015pba,Nagata:2016vkn},
and various 
techniques~\cite{Aarts:2009dg,Seiler:2012wz,Ito:2016efb,Attanasio:2018rtq} 
have been developed
to enable stable simulations satisfying these conditions.
Thanks to these developments, the CLM has been applied
successfully to finite density QCD in the heavy dense 
limit~\cite{Seiler:2012wz,Aarts:2013uxa,Aarts:2016qrv}
and at high temperature~\cite{Sexty:2013ica,Fodor:2015doa}.

In this work, we focus on the low temperature region 
with reasonably small quark mass.
Studies in this direction have been done so far 
by using the staggered fermions with 
two \cite{Kogut:2019qmi,Sinclair:2019ysx} or
four \cite{Nagata:2018mkb, Ito:2018jpo} flavors
and by using the Wilson fermions~\cite{Scherzer:2019weu}.
In our previous work using the staggered fermions with four flavors,
we employed an $8^{3}\times 16$ lattice~\cite{Ito:2018jpo},
where we have found that the CLM can be justified 
even at large quark chemical potential $\mu/T \lesssim 8$,
without using the deformation technique~\cite{Ito:2016efb}.
As a criterion for justifying the CLM, 
we used the probability distribution 
of the drift term~\cite{Nagata:2016vkn}.
Namely we consider that the results of the CLM are reliable
if the probability distribution falls off exponentially or faster.
This property of the distribution 
was indeed observed in some region of the quark chemical potential 
due to the gap 
in the Dirac eigenvalue distribution 
along the real axis \cite{Tsutsui:2019gwn}.

Within the region in which the CLM is reliable,
the baryon number density is found to exhibit 
a plateau as a function of the chemical potential.
Since this result is totally different from 
that of the phase quenched model,
the effect of the phase of the fermion determinant,
which is expected to be implemented correctly in the CLM,
must be playing an important role.
However,
we did not have
a clear physical interpretation
of this plateau behavior at that time
partly because we thought that 
the interpretation might be obscured by
severe finite volume effects due to large $\beta=5.7$,
which was chosen to stabilize the complex Langevin simulation. 
This motivated us to 
employ a $16^3\times32$ lattice,
which has double the size in each direction 
compared with the previous study.
Based on the results, we provide a clear 
interpretation of the plateau behavior.

The rest of this paper is organized as follows. 
In section~\ref{sec:2},
we briefly review how we apply the CLM to finite density
QCD and how we judge the validity of the results. 
In section~\ref{sec:Results}
we show the results obtained by the CLM on a $16^3\times32$ lattice.
In particular, we provide a clear understanding of the plateau
observed in the $\mu$-dependence of the quark number. 
The section~\ref{sec:Summary-and-discussions}
is devoted to a summary and discussions.

\section{Complex Langevin method for finite density QCD}\label{sec:2}
In this work we investigate finite density QCD with $N_{\mathrm{f}}=4$ 
flavor staggered fermions. 
After integrating out fermion fields,
the partition function reads
\begin{equation}
Z=\int \prod_{x,\nu} dU_{x,\nu}\,\det M\left[U;\mu\right]e^{-S_{\rm g}[U]} \ ,
\end{equation}
where $U_{x,\nu}\in {\rm SU}(3), \, (\nu = 1,2,3,4)$ are the link variables 
with $x = (x_1, x_2, x_3, x_4)$ being the coordinates of each site.
The action $S_{\rm g}[U]$ is defined by
\begin{align}
S_{\rm g}
= -\frac{\beta}{6}
\sum_{x}\sum_{\mu <\nu}\mathrm{tr}
\Big(U_{x,\mu\nu}+U_{x,\mu\nu}^{-1}\Big) \ , 
\quad
U_{x,\mu\nu} = U_{x\mu}U_{x+\hat\mu,\nu}U_{x+\hat\nu,\mu}^{-1}U_{x\nu}^{-1} \ .
\end{align}
Since the fermion determinant $\det M[U,\mu]$ becomes complex 
for nonzero chemical potential,
standard Monte Carlo methods suffer from the sign problem.
 
In order to overcome this problem, we apply the CLM,
which is a complex extension of the stochastic quantization 
based on the Langevin equation. 
In this method, 
the link variables $U_{x,\nu}$ are
complexified as $\mathcal{U}_{x,\nu}\in {\rm SL}(3,\mathbb{C})$, and 
accordingly
the drift term and the observables, which are functions of $U_{x,\nu}$,
have to be extended to functions of $\mathcal{U}_{x,\nu}$ 
holomorphically.
The complexified link variables are updated according to the complex version
of the Langevin equation
\begin{equation}
\mathcal{U}_{x,\nu}(t+\epsilon)=
\exp\left[i\left(-\epsilon v_{x,\nu}(\mathcal{U}(t))
+\sqrt{\epsilon}\eta_{x,\nu}(t)\right)\right]
\mathcal{U}_{x,\nu}(t) \ ,
\label{eq:cle}
\end{equation}
where $t$ is the discretized Langevin time
and $\epsilon$ is the stepsize. 
The drift term $v_{x,\nu}(\mathcal{U})$ in eq.~(\ref{eq:cle})
is defined by the holomorphic extension of
\begin{align}
v_{x,\nu}(U)
= \sum_a  \lambda_a
\left.\frac{d}{d\alpha}
S(e^{i \alpha \lambda_a}U_{x ,\nu})\right|_{\alpha=0}  \ ,
\label{drift}
\end{align}
where 
$S[U]=S_{\rm g}[U]-\ln\det M[U;\mu]$
and 
$\lambda_a \ (a=1,\cdots,8)$ are the generators of SU(3)
normalized by $\mathrm{tr}(\lambda_a \lambda_b) = \delta_{ab}$.
The noise term $\eta_{x,\nu}(t)$ in eq.~(\ref{eq:cle}), which
are $3\times3$ traceless Hermitian matrices, are generated with the
Gaussian distribution 
$\exp \left(-\frac{1}{4}\mathrm{tr}\right\{\eta_{x,\nu}^{2}(t)\left\}\right)$.

The expectation value of the observable $O(U)$
can be obtained as
\begin{equation}
\langle O(U)\rangle=
\lim_{T\rightarrow\infty}\frac{1}{T}\int_{t_{0}}^{t_{0}+T}dt\,
\langle O(\mathcal{U}(t)) \rangle_{\eta} \ ,
\end{equation}
where the expectation value 
$\langle \ \cdot  \  \rangle_{\eta}$ 
on the right-hand side
should be taken with respect to the
Gaussian noise $\eta$, 
and $t_{0}$ should be sufficiently large to achieve thermalization. 
The effect of the complex fermion determinant is
supposed to be included through the complex drift term
in the above process (\ref{eq:cle}),
and there is no need for reweighting unlike 
the path-deformation based approach such as 
the generalized Lefschetz-thimble method.
The observable we focus on here is the quark number
\begin{align}
N_\mathrm{q}=\frac{1}{L_\mathrm{t}} \frac{\partial}{\partial\mu}\ln Z \ ,
\label{def-n}
\end{align}
where $L_\mathrm{t}$ is the number of sites in the temporal direction,
which gives the inverse temperature in units of the lattice spacing.
%

In order to judge whether the obtained results are correct or not,
we use the criterion proposed in Ref.~\cite{Nagata:2016vkn}.
Let us define the magnitude of the drift term as
\begin{equation}
u= \sqrt{\frac{1}{3} 
	\max_{x, \nu}\mathrm{tr}(v_{x,\nu}^{\dagger} v_{x,\nu})} \ ,
\label{eq:drift_norm}
\end{equation}
and consider its probability distribution $p(u)$.
If $p(u)$ falls off exponentially or faster, the result is reliable. 
Conversely, the CLM is not justified when $p(u)$ shows a power law fall-off.
One of the origins of the power law fall-off is referred to as
the singular-drift problem~\cite{Nishimura:2015pba}, 
which occurs in our case when the Dirac operator acquires
near-zero eigenvalues frequently.
Another origin is referred to as the excursion problem~\cite{Aarts:2009uq}, 
which occurs when the unitarity norm
\begin{equation}
{\cal N}=\frac{1}{12 N_{\rm V}}\sum_{x,\nu}
\mathrm{tr}(\mathcal{U}_{x,\nu}^{\dagger} \mathcal{U}_{x,\nu}-{\bf 1}) 
\end{equation}
becomes too large, where $N_{\rm V}$ is the number of lattice sites.
In order to avoid this problem,
we perform the gauge cooling \cite{Seiler:2012wz}. 
(See Ref.~\cite{Nagata:2016vkn} for justification.)
Namely we perform a complexified gauge transformation
\begin{equation}
\mathcal{U}_{x,\nu}\rightarrow 
g_{x}\mathcal{U}_{x,\nu}g_{x+\hat{\nu}}^{-1} \ , \quad
\mathrm{where~} g_{x}\in {\rm SL}(3,C) 
\end{equation}
in such a way that the unitarity norm is minimized
after updating $\mathcal{U}_{x,\nu}$ by the complex Langevin equation
(\ref{eq:cle}).
We have confirmed that the excursion problem does not occur
for the parameter region explored in this work.

\section{Results\label{sec:Results}}
We have performed simulations on a $16^{3}\times32$ lattice with $\beta=5.7$
and the quark mass $m=0.01$. 
The Langevin stepsize is chosen initially as $\epsilon = 10^{-4}$
and is reduced adaptively when the magnitude of the drift exceeds 
a certain threshold~\cite{Aarts:2009dg}.
We have made $(5 \sim 15 ) \times 10^{5}$ total Langevin steps 
for each set of parameters.

\begin{figure}[bth]
\centering
\includegraphics[width=10cm]{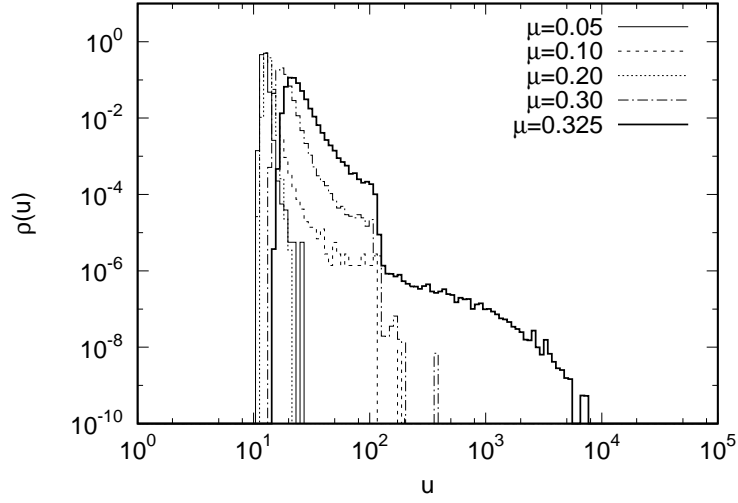}
\caption{The probability distributions of the drift term 
for various values of the chemical potential.
		\label{fig:drift}
}
\end{figure}

Let us first check the validity of the CLM.
In Fig.~\ref{fig:drift}, 
the probability distribution of the drift term $p(u)$ is 
plotted\footnote{Due to a bug
found recently in our parallel code, the magnitude of the drift
is calculated by $u = \sqrt{ \frac{1}{3} 
\mathrm{tr}(v_{x,\nu}^{\dagger} v_{x,\nu}) }$
for a randomly chosen $(x ,\nu)$
instead of using (\ref{eq:drift_norm}) with the maximum.
We have found that this bug affects the distribution of the drift term
by shifting it slightly to left, but the qualitative behavior of the tail 
seems to be unaltered. We will address this issue in the full paper.} 
for various values of $\mu$.
We find for $\mu \leq 0.3$ that $p(u)$ shows a clear exponential
fall-off and hence the simulations are reliable. 
For $\mu > 0.3$, on the other hand, the distribution
$p(u)$ shows a power law fall-off, 
suggesting that the singular drift problem occurs.

\begin{figure}[bth]
	\begin{centering}
		\includegraphics[width=9.5cm]{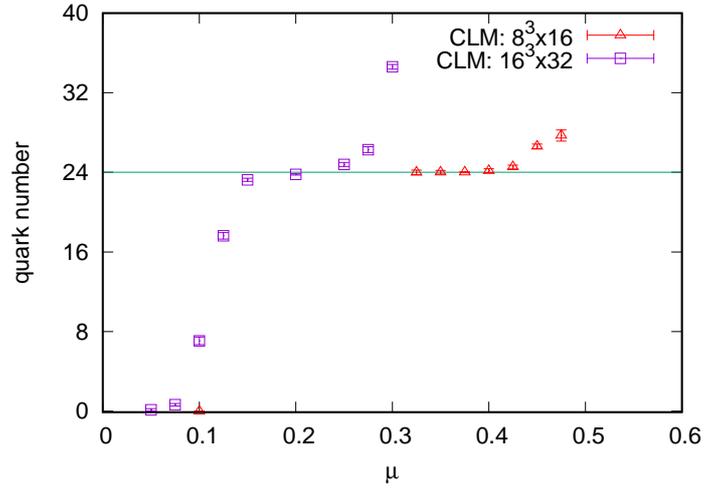}
		\caption{The quark number is plotted against
the quark chemical potential $\mu$. 
			The triangles and squares represent 
the CLM results on $8^3 \times 16$ and $16^3 \times 32$ 
			lattices, respectively. 
We show only the data that are reliable
judging from the probability distribution of the drift term.
			\label{fig:baryon}
		}
		\par\end{centering}
\end{figure}

In Fig.~\ref{fig:baryon},
we plot the quark number
against the quark chemical potential $\mu$,
where we also plot the previous results 
obtained on a $8^3 \times 16$ lattice \cite{Ito:2018jpo} for comparison.
Here we plot only the data that are reliable
judging from the probability distribution of the drift term.
Surprisingly, for both lattice sizes,
we observe a plateau 
at the height of $N_\text{q}=24$,
although the plateau region is shifted towards smaller values of $\mu$
for the larger lattice.

Here we provide a physical interpretation of this behavior.
The first thing to note is that 
the spatial size of our lattice is as small as
$L_\text{s} a \simeq$ 0.36 and 0.73 fm 
for $L_\text{s} = 8$ and 16, respectively,
since the lattice spacing $a$ is estimated as $a\simeq 0.045$ fm 
for $\beta=5.7$ and $m=0.01$ with $N_{\mathrm{f}}=4$ staggered fermions.
Therefore, the effective gauge coupling is actually small due to
the asymptotic freedom, which makes the free fermion picture valid.
The energy of a quark is given by $E({\bf p})=\sqrt{{\bf p}^2 + m^2}$
for the discrete momentum ${\bf p}= \frac{2\pi}{L_\text{s}} {\bf n}$, where
${\bf n}$ is a three-dimensional integer vector considering the periodic 
boundary conditions imposed on the spatial directions.
At zero temperature and for the quark chemical potential
within the region $m \le \mu \le \sqrt{(\frac{2\pi}{L_\text{s}})^2 + m^2}$,
the path integral should be dominated by
a state with the maximum number of quarks 
with zero momentum
allowed by the Pauli principle.
This number is $4 \times 3 \times 2 =24$, considering the
number of flavor, color and spin degrees of freedom.
Our results can be qualitatively understood in this way.

In order to have quantitative understanding of our results,
we first need to take into account the finite temperature effects
considering that the temperature used in our simulation
is about $T \simeq$ 270 and 140 MeV 
for $N_\text{t} = $ 16 and 32, respectively.
These energy scales are higher than or comparable with the 
critical temperature
$T_\text{c}/\sqrt{\sigma} \simeq 0.4$
for $N_{\mathrm{f}}=4$ staggered fermions~\cite{Engels:1996ag},
where $\sigma$ is the string tension.
In the forthcoming paper \cite{work_in_prog},
we show that our results can be understood semi-quantitatively
by the $\mu$-dependence of the quark number that
can be obtained by using the Fermi distribution function 
for finite temperature.
(See Ref.~\cite{Matsuoka:1983wy} for related work.)
Furthermore, the remaining discrepancies can be understood 
as the effects of the interactions,
which is confirmed by calculating the quark number
using the lattice perturbation theory.

\section{Summary and discussions\label{sec:Summary-and-discussions}}
In this work we have investigated finite density QCD
by the CLM
in the low temperature region with reasonably small quark mass
using four-flavor staggered fermions on a $16^3 \times 32$ lattice.
In particular, we have found that 
the quark number exhibits a plateau as a function of the chemical potential.
In the plateau region, the quark number turns out to be 24, 
which is exactly the same as what was found in our previous work 
on an $8^3 \times 16$ lattice
although the plateau region shifts towards smaller values of $\mu$.
We have provided a qualitative understanding of these behaviors
based on the free fermion picture at zero temperature.
The plateau appears because of the gap in the energy spectrum
of fermions due to the finite volume.
As the quark chemical potential is increased, only the fermions with
zero momentum can be generated.
The number 24, which appears as the height of the plateau
in the plot of the quark number
can be understood as the number of internal degrees of freedom
of the fermion. 
The quark number is expected to jump to the second plateau
at $\mu \sim \frac{2\pi}{L_\text{s}}$.
While this behavior is somewhat obscured by finite temperature effects
in our setup, the shift of the plateau region
can be understood naturally.

The significance of this result is two-fold.
First, the fact that the behaviors observed by the CLM
have found a clear physical interpretation confirms further that
the method is working and we are getting correct results.
While the interpretation shows that we are well within the perturbative
regime due to the small spatial extent and the asymptotic freedom,
it would be nice to test the method nevertheless in a region in which
we can obtain explicit results perturbatively.
To our knowledge, the behaviors reported above have never been observed
by other methods, which clearly demonstrates 
the usefulness of the CLM.
Second, the condensation of the zero momentum modes may be regarded
as the beginning of the formation of the Fermi sphere, which is crucial
in color superconductivity. By increasing the lattice size further,
the Fermi sphere is expected to 
grow due to condensation of higher momentum modes.
If such behaviors can be observed within the applicability of the CLM,
we should be able to study the color superconductivity 
in the near future.

\section*{Acknowledgements}
We would like to thank A.~Onishi, M.~Scherzer for important discussions
and M.P.~Lombardo for drawing our attention
to Ref.~\cite{Matsuoka:1983wy},
which led us to the present interpretation of our results.
The authors are also grateful to Y.~Asano, 
Y.~Namekawa and T.~Yokota for fruitful discussions.
This research was supported by MEXT as
``Priority Issue on Post-K computer'' 
(Elucidation of the Fundamental Laws and Evolution of the Universe) 
and Joint Institute for Computational Fundamental Science (JICFuS).
Computations were carried out using computational resources of the K computer 
provided by the RIKEN Advanced Institute for Computational Science 
through the HPCI System Research project (Project ID:hp180178, hp190159).
S.~T. was supported by the RIKEN Special Postdoctoral Researchers Program.
J.~N.\ was supported in part by Grant-in-Aid 
for Scientific Research (No.\ 16H03988)
from Japan Society for the Promotion of Science. 
S.~S.\ was supported by the MEXT-Supported Program 
for the Strategic Research Foundation at Private Universities 
``Topological Science'' (Grant No.\ S1511006).

\bibliography{lat2019.bib}
\bibliographystyle{h-physrev5}

\end{document}